\pdfoutput=1
\documentclass[journal=jctcce,manuscript=article,layout=traditional]{achemso}
\setkeys{acs}{%
  super=true, %
  articletitle=true, %
  doi=false, %
  etalmode=truncate, %
  maxauthors=0 %
}

\usepackage{amsmath,amsfonts,amssymb} %
\usepackage{mathtools} %
\ifx\phone\undefined
\usepackage{wasysym} 
\else
\let\Phone\phone
\let\phone\relax
\usepackage{wasysym}

\let\phone\Phone
\let\Phone\relax
\fi
\usepackage{bm} %
\usepackage[bbgreekl]{mathbbol} %
\usepackage{graphicx} %
\usepackage[dvipsnames,table]{xcolor} %
\usepackage{multirow,tabularx} %
\usepackage[acronym]{glossaries} %
\usepackage{braket} %
\usepackage[binary-units=true,detect-weight=true]{siunitx} %
\usepackage{fancyhdr} %
\usepackage{microtype} %
\usepackage{natmove} 
\usepackage[caption=false]{subfig} %
\usepackage[byname]{smartref} %
\usepackage[breaklinks, %
colorlinks, linkcolor=Blue, citecolor=Blue, urlcolor=Blue]{hyperref} %
\usepackage[version=3]{mhchem} %
\usepackage{wrapfig} %
\AtBeginDocument{\usepackage{booktabs}} 
\usepackage[linesnumbered,lined,boxed,ruled]{algorithm2e}

\DeclareSIUnit[number-unit-product = {\,}]\cal{cal}
\def\MyTitle{Efficient construction of involutory linear combinations
  of anti-commuting Pauli generators for large-scale iterative qubit
  coupled cluster calculations} %
\def\MyAuthora{Ilya G. Ryabinkin} %
\def\MyAuthorb{Andrew J. Jena} %
\def\MyAuthorc{Scott N. Genin} %

\def\MySubject{} %

\hypersetup{%
  pdftitle={\MyTitle{}}, %
  pdfauthor={\MyAuthora{}, \and \MyAuthorb{}, \and \MyAuthorc{}}, %
  pdfsubject={\MySubject{}}, %
  pdfkeywords={} %
}

\graphicspath{{pics/}}

\newcolumntype{Y}{>{\centering\arraybackslash}X}

\newacronym[longplural={degrees of freedom}, firstplural={degrees of
  freedom (DOF)}, plural={DOF}]{DOF}{DOF}{degree of freedom} %
\newacronym[longplural={equations of motion}, firstplural={equations
  of motion (EOM)}, plural={EOM}]{EOM}{EOM}{equation of motion} %

\newacronym{OLED}{OLED}{organic light-emitting diode}
\newacronym{NISQ}{NISQ}{noisy intermediate-scale quantum}

\newacronym{JW}{JW}{Jordan--Wigner} %
\newacronym{BK}{BK}{Bravyi--Kitaev} %

\newacronym{QPE}{QPE}{quantum phase estimation} %
\newacronym{VQE}{VQE}{variational quantum eigensolver} %
\newacronym{QMF}{QMF}{qubit mean-field} %
\newacronym{QCC}{QCC}{qubit coupled cluster} %
\newacronym{iQCC}{iQCC}{iterative qubit coupled cluster} %
\newacronym{PQA}{PQA}{parametrized quantum annealing} %
\newacronym{DIS}{DIS}{direct interaction set} %
\newacronym[longplural={involutary linear combinations of
anti-commuting Paulis}, firstplural={involutory linear combinations of
anti-commuting Paulis (ILCAP)}, plural={ILCAP}]{ILCAP}{ILCAP}{involutory linear combination of anti-commuting Paulis} %

\newacronym{CAS}{CAS}{complete active space} %
\newacronym{PES}{PES}{potential energy surface} %
\newacronym{PEC}{PEC}{potential energy curve} %
\newacronym{AO}{AO}{atomic orbital} %
\newacronym{MO}{MO}{molecular orbital} %

\newacronym{CI}{CI}{configuration interaction} %
\newacronym{FCI}{FCI}{full configurational interaction} %
\newacronym{CASCI}{CASCI}{complete active space configurational
  interaction} %
\newacronym{MCSCF}{MCSCF}{multiconfigurational self-consistent
  field} %
\newacronym{CASSCF}{CASSCF}{complete active space self-consistent
  field} %
\newacronym{CC}{CC}{coupled cluster} %
\newacronym{UCC}{UCC}{unitary coupled cluster} %
\newacronym{UCCSD}{UCCSD}{unitary coupled cluster singles and
  doubles} %
\newacronym{CCSD}{CCSD}{coupled-cluster singles and doubles} %
\newacronym{CCSD-T}{CCSD(T)}{coupled-cluster singles and doubles and
  non-iterative triples} %
\newacronym{RHF}{RHF}{restricted Hartree--Fock} %
\newacronym{CIS}{CIS}{configuration-interaction singles} %
\newacronym{ROHF}{ROHF}{restricted open-shell Hartree--Fock} %
\newacronym{UHF}{UHF}{unrestricted Hartree--Fock} %
\newacronym{DMRG}{DMRG}{density-matrix renormalization group} %
\newacronym{DFT}{DFT}{density-functional theory} %
\newacronym{TDDFT}{TDDFT}{time-dependent density-functional theory} %
\newacronym{ENPT}{ENPT}{Epstein-Nesbet perturbation theory} %
\newacronym{MP}{MP}{M{\o}ller--Plesset perturbation theory} %
\newacronym{MP2}{MP2}{second-order M{\o}ller--Plesset perturbation theory} %
\newacronym{MRMP2}{MRMP2}{second-order multi-reference M{\o}ller--Plesset perturbation theory} %

\newacronym{SQP}{SQP}{sequential quadratic programming} %
\newacronym{MMA}{MMA}{method of moving asymptotes} %

\newcommand{\E}{\textrm{e}} %
\newcommand{\I}{\mathrm{i}\mkern1mu} %
\def\be{\begin{equation}} %
\def\ee{\end{equation}} %
\def\bea{\begin{eqnarray}} %
\def\eea{\end{eqnarray}} %

\title{\MyTitle}

\author{\MyAuthora{}} %
\affiliation{OTI Lumionics Inc., 3415 American Drive Unit 1, \\
  Mississauga, Ontario L4V\,1T4, Canada} %
\email{ilya.ryabinkin@otilumionics.com} %

\author{\MyAuthorb{}} %
\affiliation{Combinatorics \& Optimization, University of Waterloo \\
  Waterloo, Ontario, N2L\,3G1, Canada} %

\author{\MyAuthorc{}} %
\affiliation{OTI Lumionics Inc., 3415 American Drive Unit 1, \\
  Mississauga, Ontario L4V\,1T4, Canada} %

\begin{document}
\ifx\latin\undefined %
  \newcommand{\latin}{\textit} %
\fi
\newcommand{\gate}[1]{\ensuremath{\mathtt{#1}}}

\date{\today}

\begin{abstract}
  We present an efficient method for construction of a fully
  anti-commutative set of Pauli generators (elements of the Pauli
  group) from a commutative set of operators that are composed
  exclusively from Pauli $\hat x_i$ operators (purely X~generators)
  and sorted by an associated numerical measure, such as absolute
  energy gradients. Our approach uses the Gauss--Jordan elimination
  applied to a binary matrix that encodes the set of X~generators to
  bring it to the reduced row echelon form, followed by the
  construction of an anti-commutative system in a standard basis by
  means of a modified Jordan-Wigner transformation and returning to
  the original basis. The algorithm complexity is linear in the size
  of the X~set and quadratic in the number of qubits. The resulting
  anti-commutative sets are used to construct the qubit coupled
  cluster Ansatz with involutory linear combinations of anti-commuting
  Paulis (QCC-ILCAP) proposed in [J. Chem. Theory
  Comput.~\textbf{2021}, \textit{17}, 1, 66–78]. We applied the
  iterative qubit coupled cluster method with the QCC-ILCAP Ansatz to
  calculations of ground-state potential energy curves for symmetric
  stretching of the water molecule (36 qubits) and dissociation of
  \ce{N2} (56 qubits).
\end{abstract}

\glsresetall

\maketitle

\section{Introduction}
\label{sec:introduction}

Quantum chemistry calculations are among the most promising
applications of quantum computing. A lot of efforts have been spent
towards quantum computer-friendly algorithms for solving the
electronic structure problem in particular~\cite{Cao:2019/cr/10856,
  Bauer:2020/cr/12685, McArdle:2020/rmp/015003,
  Motta:2022/wires/e1580, Tilly:2022/arXiv/2111.05176}. The major
obstacle in designing such algorithms is that the current and
near-term quantum computers are the
\gls{NISQ}~\cite{Preskill:2018/quant/79} devices featuring limited
number of qubits, limited connectivity, short coherence times and high
levels of noise. To cope with these limitations, the \gls{VQE}
approach has been proposed~\cite{Peruzzo:2014/ncomm/4213}. In
\gls{VQE}-based methods a quantum computer runs a parametrized quantum
circuit to prepare a trial quantum state, then Hamiltonian terms are
repeatedly measured individually or in
groups~\cite{Poulin:2018/prl/010501,Huggins:2019/arXiv/1907.13117,
  Crawford:2019/arXiv/1908.06942, Verteletskyi:2020/jcp/124114,
  Izmaylov:2019/cs/3746, Yen:2020/jctc/2400} on this state to obtain
the ground-state energy estimate. Subsequently, accumulated estimates
are used by a classical computer to predict a location of the energy
minimum via gradient-free optimization; updated parameters are
returned back to a quantum computer thus forming a quantum-classical
feedback loop. Recently, efficient algorithms for measuring energy
gradients on a quantum computer have also been
proposed~\cite{Schuld:2019/pra/032331, Izmaylov:2021/pra/062443}
allowing for the use of gradient-based minimization
schemes~\cite{Piskor:2022/pra/062415}.

The quantum circuit represents a unitary transformation
${\hat U}(\mathbf{t})$ of a reference state $\ket{0}$ into a target
state $\ket{\Psi(\mathbf{t})}$ for a set of user-controlled parameters
$\mathbf{t} = t_1, \ldots, t_L$:
\begin{equation}
  \label{eq:VQE_U}
  \ket{\Psi(\mathbf{t})} = {\hat U}(\mathbf{t}) \ket{0}.
\end{equation}
The unitary transformation must be realizable on a quantum computer --
in other words, it must be readily translated into a sequence of
quantum gates without additional approximations. This can be trivially
achieved if ${\hat U}(\mathbf{t})$ is directly encoded in terms of
gates available on particular quantum
hardware~\cite{Kandala:2017/nature/242} but a far more popular
approach is to employ some universal, hardware-independent
intermediate representation. As such, a product of exponents of Pauli
generators $\hat T_k$ is used:
\begin{equation}
  \label{eq:U_QCC}
  \hat U(\mathbf{t}) = \prod_{k=1}^L\exp\left(-\I t_k \hat T_k/2\right).
\end{equation}
Generators (``Pauli words'')
\begin{equation}
  \label{eq:P_k-def}
  \hat T_k = \prod_{j \ge 0,\ j \in j(k)} \hat \sigma_j,
\end{equation}
are strings (tensor products) of Pauli elementary operators
$\hat \sigma_j = \hat x_j$, $\hat y_j$, or $\hat z_j$, where
$0 \le j \le (n-1)$ and $n$ is the number of qubits. There are
$4^n - 1$ non-trivial Pauli words for $n$ qubits; together with the
identity operator $\mathcal{\hat I}$ and phase factors $\pm 1,\ \pm\I$
they constitute $4^{n+1}$-element Pauli group with respect to
multiplication~\cite[chap~10.5.1]{Nielsen:2010}.

Equation~\eqref{eq:U_QCC} is the final form for many \gls{VQE}-based
methods~\cite{Ortiz:2001/pra/022319, Peruzzo:2014/ncomm/4213,
  Wecker:2015/pra/042303, Mcclean:2016/njp/023023,
  OMalley:2016/prx/031007, Romero:2018/qct/014008,
  Ryabinkin:2018/jctc/6317, Grimsley:2018/nc/3007, Lee:2019/jctc/311,
  Nam:2020/npj_qi/33, Ryabinkin:2020/jctc/1055}. They differ, however,
in the way how generators $\hat T_k$ are selected and ordered. The
\gls{QCC} method~\cite{Ryabinkin:2018/jctc/6317} constructs the
Ansatz~\eqref{eq:U_QCC} directly by selecting appropriate elements of
Pauli group based on the energy gradient criterion. As we have
shown~\cite{Ryabinkin:2020/jctc/1055}, there exists a linearly scaling
(in the number of Hamiltonian terms) algorithm that allows one to rank
Pauli words in their projected importance for the energy lowering.
Paired with the iterative approach, this ranking scheme constitutes a
basis for the \gls{iQCC} method, which has been subsequently augmented
with perturbative completeness corrections to treat large (more than
70 qubits) systems~\cite{Ryabinkin:2021/qst/024012,
  Genin:2022/ang/e202116175}.

Besides suitability for quantum computers, Eq.~\eqref{eq:U_QCC} is
straightforward to implement on classical computers. Because Pauli
words with unit phases are involutory operators,
\begin{equation}
  \label{eq:involutory_prop}
  \hat T_k^2 = \mathcal{\hat I},\ \forall k,
\end{equation}
their exponentiation is trivial:
\begin{equation}
  \label{eq:expT}
  \exp\left(-\frac{\I t_k\hat T_k}{2}\right) = \cos\left(\frac{t_k}{2}\right) -
  \I\sin\left(\frac{t_k}{2}\right)\hat T_k.
\end{equation}
Plugging Eq.~\eqref{eq:expT} into Eq.~\eqref{eq:U_QCC} and expanding,
we obtain a sum of $2^L$ terms
\begin{equation}
  \label{eq:U_QCC_expanded}
  \hat U(\mathbf{t}) = \prod_{k=1}^L\cos\left(\frac{t_k}{2}\right) -\I
  \sum_{j=1}^L \hat T_j \sin\left(\frac{t_j}{2}\right)
  \prod_{k\ne j}^L\cos\left(\frac{t_k}{2}\right) - \dots
\end{equation}
thus illustrating the exponential complexity of the Ansatz. While this
complexity is not a problem for a perfect quantum computer---and can
be even considered as quantum advantage---it is an obstacle for
classical simulators and for implementation of various pre- and
post-processing techniques that decrease the amount of work performed
by a \gls{NISQ} device.

An Ansatz that is characterized by linear, not exponential, complexity
has been proposed in Ref.~\citenum{Lang:2021/jctc/66}. It is based on
a few simple ideas. The closed form of the exponent of generators,
Eq.~\eqref{eq:expT} exists because of the involutory property of Pauli
words, Eq.~\eqref{eq:involutory_prop}. Hence, one can try to
generalize this property to linear combinations of Pauli words,
namely, if
\begin{eqnarray}
  \label{eq:T_lc}
  \hat T & = & \sum_{k=1}^M \alpha_k \hat T_k, \\
  \label{eq:T_lc_involutory}
  \hat T^2 & = & \mathcal{\hat I},
\end{eqnarray}
then
\begin{equation}
  \label{eq:U_ilcap}
  \begin{split}
    \hat U(t, \boldsymbol \alpha) & = \exp\left(-\frac{\I t\hat
                                    T}{2}\right) = \cos\left(\frac{t}{2}\right) -
                                    \I\sin\left(\frac{t}{2}\right)\hat T \\
                                  & = \cos\left(\frac{t}{2}\right) -
                                    \I\sin\left(\frac{t}{2}\right)\sum_{k=1}^M \alpha_k \hat T_k.
  \end{split}
\end{equation}
Performing elementary algebraic manipulations with
Eqs.~\eqref{eq:T_lc} and \eqref{eq:T_lc_involutory} we obtain:
\begin{eqnarray}
  \label{eq:alpha_cond}
  \sum_{k=1}^M \alpha_k^2 & = & 1, \\
  \label{eq:antisymm_cond}
  \left[\hat T_k, \hat T_m\right]_{+} & = & 0, \quad 1 \le k,m \le M,\ k
                                            \ne m
\end{eqnarray}
where $[\hat A, \hat B]_{+} = \hat A \hat B + \hat B \hat A$ is the
anti-commutator. Thus, the vector of coefficients $\boldsymbol\alpha$
must be normalized, and Pauli words must be all anti-commuting. In
general, any two Pauli words are either commuting or anti-commuting,
hence, there exist some non-trivial (\latin{i.e.} with $M \ge 2$)
linear combinations of Paulis that satisfy
Eq.~\eqref{eq:antisymm_cond}. The main result of
Ref.~\citenum{Lang:2021/jctc/66} is that $M \le 2n-1$ where $n$ is the
number of qubits when $\hat T_k$ are subjected to the gradient
condition, which \emph{guarantees} the energy lowering in \gls{VQE}
optimization to avoid the so-called \textit{barren
  plateaus}~\cite{Mcclean:2018/nc/4812}.
Reference~\citenum{Lang:2021/jctc/66} has also presented a
proof-of-the-principle algorithm for constructing such \glspl{ILCAP}.

The current paper builds upon and extends
Ref.~\citenum{Lang:2021/jctc/66}. First of all, we propose a new
algorithm for constructing the anti-commuting sets of Pauli operators
from a set of those that are ``tagged'' by an additional property,
such as the absolute energy gradient (see
Sec.~\ref{sec:outl-glsiqcc-meth}). Our algorithm makes use of
Gauss--Jordan elimination for matrices with coefficients from the
Galois field GF(2) -- the binary numbers 0 and 1. It is highly
efficient and allows for dealing with problems with the number of
qubits $n \gtrsim 100$ and Hamiltonians containing billions of terms.
Contrary to the original algorithm from
Ref.~\citenum{Lang:2021/jctc/66} it does not require any
trial-and-error steps. Finally, the new algorithm is capable of
attaining the upper limit for the size of anti-commuting sets,
$M = 2n-1$ though practical considerations may suggest lower $M$.

The paper is organized as follows. First, we briefly outline the
\gls{iQCC} method. Secondly, we introduce a new algorithm for
constructing systems of anti-commuting generators. Then we discuss a
new variant of the perturbation-theory correction to the
QCC-\gls{ILCAP} Ansatz and compare it to the perturbative correction
proposed in
Ref.~\citenum{Ryabinkin:2021/qst/024012}. 
Finally, we assess the performance of the QCC-\gls{ILCAP} Ansatz as a
pre- and post-processing technique within the ``standard'' \gls{iQCC}
method on cases of large-scale CI calculations for symmetric
dissociation of the \ce{H_2O} molecule and stretching of the \ce{N2}
molecule.

\section{Theory}
\label{sec:theory}

\subsection{An outline of the \gls{iQCC} method}
\label{sec:outl-glsiqcc-meth}

The \gls{iQCC} method takes an active-space second-quantized
electronic Hamiltonian of a molecule~\cite{Helgaker:2000,
  Abrams:1997/prl/2586, AspuruGuzik:2005/sci/1704,
  Peruzzo:2014/ncomm/4213},
\begin{equation}
  \label{eq:qe_ham}
  \hat H_e = E_\text{core} + \sum_{ij} f_{ij} {\hat a}^\dagger_i {\hat
    a}_j + \frac{1}{2}\sum_{ijkl}
  g_{ijkl} {\hat a}^\dagger_i {\hat a}^\dagger_j {\hat a}_l {\hat a}_k,
\end{equation}
as input. ${\hat a_i}^\dagger$ and ${\hat a_i}$ are fermion creation
and annihilation operators in the active space, $E_\text{core}$ is the
electronic energy associated with inactive (core) orbitals, $f_{ij}$,
and $g_{ijkl}$
are one- and two-electron contributions to the electronic energy
written in a spin-orbital basis.
\begin{align}
  \label{eq:2eints}
  g_{ijkl} = & \iint
               \frac{\psi_i^{*}(\mathbf{x}_1)\psi_k^{*}(\mathbf{x}_2)\psi_j(\mathbf{x}_1)\psi_l(\mathbf{x}_2)}{r_{12}}
               \mathrm{d}\mathbf{x}_1 \mathrm{d}\mathbf{x}_2 
\end{align}
where $\mathbf{x} = (\mathbf{r}, \sigma)$. $f_{ij}$ account for the
kinetic and nuclear-attraction energy of active electrons as well as
their electrostatic interaction with core electrons.

Prior to any computations, the Hamiltonian~\eqref{eq:qe_ham} is
converted to a qubit form by the \gls{JW}
transformation~\cite{Jordan:1928/zphys/631,AspuruGuzik:2005/sci/1704}
to obtain
\begin{equation}
  \label{eq:qubit_H}
  \hat H = \sum_{k=1}^{M} C_k \hat P_k,
\end{equation}
where $C_k$ are coefficients inferred from $f_{ij}$ and $g_{ijkl}$ and
$\hat P_k$ are Pauli words.

Apart from the Hamiltonian, the \gls{iQCC} method requires a reference
vector $\ket{0}$ to be specified. As such, a direct-product state,
\begin{align}
  \label{eq:qubit_ref}
  \ket{0} & = \prod_{k=1}^{n_e} \ket{\downarrow}_k \times \prod_{k=1}
            ^{n-n_e} \ket{\uparrow}_k
\end{align}
is used. Here $n_e$ is the number of electrons in the active space for
an electronic state of interest and $n$ is the total number of qubits
which is equal to the number of active spin-orbitals. If $n$ is twice
the size of an atomic basis and $n_e$ equals to the total number of
electrons in a molecule (hence, $E_\text{core} = 0$), one deals with
the \gls{FCI} problem, otherwise it is a \gls{CASCI} problem with an
active space which is commonly abbreviated as CAS($n_e$, $n/2$).
$\ket{\uparrow}_k$ and $\ket{\downarrow}_k$ are eigenstates of
$\hat z_k$ with eigenvalues $+1$ and $-1$, respectively. We
additionally assume pairwise grouping of spin-orbitals: the first
orbital with $\alpha$ (``up'') spin is mapped to the first qubit,
followed by the first orbital with $\beta$ spin, which is mapped to
the second, etc. There is one-to-one correspondence between
spin-orbital population strings for fermions and product states of
qubits if the \gls{JW} transformation is employed. Thus, if the basis
of Hartree--Fock \glspl{MO} is used, $\ket{0}$ represents an electron
configuration that satisfies the Aufbau principle, which is typically
the lowest-energy configuration.

An essential feature of the \gls{iQCC} method is how generators
$\hat T_k$ are selected and ordered (``ranked'') to be used in the
Ansatz~\eqref{eq:U_QCC}. For the sake of brevity we introduce the
simplest gradient-based ranking scheme; other possibilities are
discussed in Ref.~\citenum{Ryabinkin:2021/qst/024012}. Consider a
single-generator \gls{QCC} Ansatz, $\hat U(t) = \exp(-\I t\hat T/2)$,
and compute the energy expectation value for
$\ket{\Psi(t)} = \hat U(t) \ket{0}$:
\begin{align}
  \label{eq:E_T}
  E[\hat T](t) & = \braket{\Psi(t)|\hat H|\Psi(t)} = \braket{0|\exp(\I
                 t\hat T/2) \hat H\exp(-\I t\hat T/2)|0}. 
\end{align}
Using Eq.~\eqref{eq:expT} one can further elaborate the expression for
$E[\hat T](t)$:
\begin{equation}
  \label{eq:E_T_expanded}
  E[\hat T](t) = \braket{0|\hat H|0} + \frac{\sin(t)}{2\I}
  \Braket{0|[\hat H, \hat T]|0} +
  \frac{1 - \cos(t)}{2} \Braket{0|(\hat T \hat H \hat T
    - \hat H)|0},  
\end{equation}
where $[\hat A, \hat B] = \hat A \hat B - \hat B \hat A$ is a
commutator. Defining
\begin{equation}
  \label{eq:omegas}
  \omega  =  \left|\frac{1}{2\I}\braket{0|[\hat H, \hat T]|0}\right| =
  \left|\Im\braket{0|\hat H \hat T|0}\right|, \\
\end{equation}
we notice that $\omega$ is an absolute value of the energy gradient at
$t=0$. Generators with non-zero $\omega$ \emph{guarantee} the energy
lowering being used in the \gls{QCC} Ansatz~\eqref{eq:U_QCC}. Thus,
one can use $\omega$ for sorting generators taking $\hat T_k$ with
larger $\omega_k$ first.

All generators with \latin{a priori} non-zero values of $\omega$ can
be efficiently constructed given the Ising decomposition of the
Hamiltonian~\cite{Ryabinkin:2020/jctc/1055,
  Ryabinkin:2021/qst/024012}, which reads
\begin{equation}
  \label{eq:Ising_decomp}
  \hat H = {\hat I}_0(\mathbf{z}) + \sum_{k >0} {\hat I_k}(\mathbf{z})
  \hat X_k,
\end{equation}
where ${\hat I_k}(\mathbf{z})$, $k=0, 1, \dots$ are qubit Hamiltonians
that are sums of Pauli words containing only Pauli elementary $\hat z$
operators (``generalized Ising Hamiltonians''). All $\hat X_k$ are the
Pauli X~strings,
\begin{equation}
  \label{eq:X-strings}
  \hat X_k = \prod_{j \ge 0,\ j \in j(k)} \hat x_j.
\end{equation}
The decomposition~\eqref{eq:Ising_decomp} does not contain Pauli
$\hat y$ operators because they are factorized as
$\hat y_j = -\I \hat z_j \hat x_j$.

As was shown in Ref.~\citenum{Ryabinkin:2020/jctc/1055}, there are
$2^{n-1}$ generators $\hat T_{k_l}$ for every \emph{single} $\hat X_k$
that appears in the decomposition~\eqref{eq:Ising_decomp}, which are
all characterized by the same value of the absolute energy gradient
determined solely by the parental $\hat X_k$:
\begin{equation}
  \label{eq:omega_k}
  \omega_k = \left|\braket{0|\hat I_k|0}\right|.
\end{equation}
They all can be obtained from $\hat X_k$ by substituting an \emph{odd}
number of elementary $\hat x_j$ operators with their $\hat y_j$
counterparts and adding an arbitrary number of $\hat z$ operators with
qubit indices that are \emph{not} in $\hat X_k$. This redundancy is
exploited in the original \gls{iQCC} method to define a ``canonical
set'' of generators: $\hat T_k$ are created by converting a
\emph{single} $\hat x_j$ with the smallest possible $j$ in every
$\hat X_k$ into $\hat y_j$. More importantly, it provides necessary
flexibility for finding $M \ge 2$ anti-commuting generators satisfying
Eq.~\eqref{eq:antisymm_cond}.

After ranking and selecting generators at the $i$-th iteration, the
\gls{iQCC} method minimizes the \gls{QCC} energy expression
\begin{equation}
  \label{eq:QCC_energy_expr}
  E^{(i)}(t_1, \dots, t_L) = \Braket{0|\left({\hat U}^{(i)}\right)^\dagger(t_1,
    \dots, t_L) \hat H^{(i)} {\hat U}^{(i)}(t_1, \dots, t_L)|0}, \quad i=1, \dots
\end{equation}
where $\hat H^{(i)}$ is the current Hamiltonian
[$\hat H^{(1)} = \hat H$ is the original electronic Hamiltonian in
qubit representation, Eq.~\eqref{eq:qubit_H}], and ${\hat U}^{(i)}$ is the
\gls{QCC} Ansatz~\eqref{eq:U_QCC} with $L$ topmost generators
constructed using the Ising decomposition of $\hat H^{(i)}$.

Once the energy is minimized, the \gls{iQCC} algorithm performs
\emph{dressing} of the current Hamiltonian $\hat H^{(i)}$. Dressing is
a unitary transformation of $\hat H^{(i)}$ using
${\hat U}^{(i)}(\mathbf{t}_\text{opt}^{(i)})$ where
$\mathbf{t}_\text{opt}^{(i)}$ are the optimized amplitudes. Below we
demonstrate the first step of this transformation, which corresponds
to $L=1$ in Eq.~\eqref{eq:U_QCC}; subsequent steps (for $L > 1$) are
performed recursively:
\begin{align}
  \label{eq:dressing}
  \hat H^{(i+1)}  & = \left({\hat U}^{(i)}\right)^\dagger(t_\text{1,opt}^{(i)})
                    \hat H^{(i)} {\hat U}^{(i)}(t_\text{1,opt}^{(i)}) \\
                  & = \hat H^{(i)} -
                    \frac{\I}{2}\sin\left(t_\text{1,opt}^{(i)}\right) [\hat
                    H^{(i)}, \hat T_1] + \frac{1 -
                    \cos(t_\text{1,opt}^{(i)})}{2}\left(\hat
                    T_1 \hat H^{(i)} \hat T_1 - \hat
                    H^{(i)}\right). 
\end{align}
If dressing is performed exactly (\latin{i.e.} without rounding), the
following identity is hold:
\begin{equation}
  \label{eq:dressing_identity}
  \braket{0|\hat H^{(i+1)}|0} = E^{(i)}(t_{1,\text{opt}}^{(i)}, \dots,
  t_{L,\text{opt}}^{(i)}).
\end{equation}
The identity~\eqref{eq:dressing_identity} allows one to complete an
\gls{iQCC} loop. One can start a new iteration taking $\hat H^{(i+1)}$
as an initial Hamiltonian while keeping the reference vector $\ket{0}$
intact. The iterative procedure then proceeds until some convergence
criteria (\latin{e.g.} largest $\omega$ is less than a threshold) are
met. The \gls{QCC} energy at the final iteration then becomes the
ground-state energy estimate. We emphasise that, contrary to the
\gls{VQE}-style methods based on the \gls{UCC} hierarchy with limited
exitation rank (\latin{e.g.} UCCSD), the \gls{iQCC} method is
\emph{exact}, in other words, it is capable of attaining the
\gls{CASCI} or \gls{FCI} energy of a system.

\subsection{Constructing a set of mutually anti-commuting generators}
\label{sec:counstr-set-mutu}

\subsubsection{A problem statement}
\label{sec:problem-statement}

Given a set of Pauli X~words $\hat X_1$,
$\hat X_2, \dots, \hat X_m \equiv \{\hat X_m\}$ that are taken from
the Ising decomposition~\eqref{eq:Ising_decomp} of the current
Hamiltonian and ordered according to their absolute gradients
$\omega_1 \ge \omega_2 \ge \dots \ge \omega_m$,
Eq.~\eqref{eq:omega_k}, find a largest subset
$\{\hat X_{m_l}\} \subseteq \{\hat X_m\}$, and construct a set of
complementary Z~words $\{\hat Z_{m_l}\}$, such that the ``phaseless''
products $\hat T_{m_l} = \hat X_{m_l} \odot \hat Z_{m_l}$ contain the
\emph{odd} number of Pauli elementary operators $\hat y$ and satisfy
Eq.~\eqref{eq:antisymm_cond}. The phaseless multiplication follows the
normal multiplication rule for the Pauli elementary operators but
ignores the resulting phases $\pm \I$ completely, that is, for
example, $\hat x \odot \hat y = \hat y \odot \hat x = \hat z$. The
condition on the odd number of $\hat y$ factors in every
$\hat T_{m_l}$ ensures that the real part of the energy gradient
$\left.\mathrm{d}E[\hat T_{m_l}]/\mathrm{d}t\right|_{t=0}$ is
non-zero.

\subsubsection{An outline of the solution}
\label{sec:an-outline-solution}

The solution is found in a few steps. First, we employ the binary
encoding~\cite{Bravyi:2017/ArXiv/1701.08213} to map Pauli words to
column vectors of the length $n$ with 0 and 1 as entries.
Consequently, the set $\{\hat X_m\}$ can be represented as a
$n \times m$ matrix $\mathbf{M}$ whose columns are the binary vectors
$X_1$, $X_2$, \latin {etc}. Second, we show that two universal qubit
transformations, namely, \gate{CNOT} and \gate{SWAP} gates, in the
binary representation become the elementary matrix transformations of
$\mathbf{M}$, row-addition and row-switching, respectively. This
correspondence allows us to apply Gauss--Jordan elimination to
$\mathbf{M}$ to bring it to the reduced row-echelon
form~\cite[p~48]{Book/Meyer2000}, $\mathbf{M}_\text{rref}$. The
columns of $\mathbf{M}_\text{rref}$ containing the leading 1 are
identified with a subset of all Pauli elementary operators
$\{\hat x_j\}_{j=0}^{(n-1)}$. Then we show how to convert elementary
Pauli $x$ operators into $(2n-1)$ \emph{anti-commuting} operators with
the odd number of $\hat y$ factors using a slightly modified
\acrlong{JW} transformation. Speaking informally, the
\gls{JW}-construction solves our problem by finding such Pauli Z~words
that make the ``standard'' basis of $\hat x_j$ anti-commutative.
Finally, we map a binary representation of those Z~words back to the
original basis to determine $\{\hat Z_{m_l}\}$ which completes our
algorithm.

\subsubsection{Binary encodings for X and Z Pauli words and a matrix
  $\mathbf{M}$}
\label{sec:binary-encoding-set}

Let us assume that the number of qubits $n$ is fixed. Then, any Pauli
X~word can be mapped into an $n$-dimensional column vector
$(\dots 1 \dots 0 \dots)^T$ by setting 1 in the position $i$ as long
as $\hat x_{i-1}$ enters the word. For example, the word
$\hat X = \hat x_0 \hat x_2$ for $n=4$ is mapped to
$(1\, 0\, 1\, 0)^T$. Similarly, any Z~word can be mapped into an
$n$-vector by applying the same rule. A \emph{general} Pauli word
$\hat P$, thus, can be represented as $2n$-dimensional vector
$(X | Z)^T$ according to the factorization
$\hat P = \hat X \odot \hat Z$~\cite{Bravyi:2017/ArXiv/1701.08213}.
Since we will be always dealing with factorized expressions, it is
sufficient to consider the $n$-dimensional binary representations for
X and Z~words separately.

Consider a sequence of Pauli words
$\{\hat X_m\} = {\hat X_1, \hat X_2, \dots}$ that are taken from the
Ising decomposition~\eqref{eq:Ising_decomp} and arranged according to
their absolute gradients $\omega_1 \ge \omega_2 \ge \dots$. The
corresponding binary vectors $X_1, X_2, \dots$ are the columns of a
matrix $\mathbf{M}$. For example, a 5-element sequence of X~words
\begin{equation}
  \label{eq:example_X_set}
  \{\hat x_0 \hat x_2,\ \hat x_1 \hat x_3,\ \hat x_0 \hat x_1 \hat
  x_2,\ \hat x_1 \hat x_2 \hat x_3,\ \hat x_0 \hat x_1 \hat x_2 \hat
  x_3\} 
\end{equation}
for $n = 4$ qubits is encoded as a $4 \times 5$ matrix
\begin{equation}
  \label{eq:example_M}
  \begin{pmatrix}
    1 & 0 & 1 & 0 & 1 \\
    0 & 1 & 1 & 1 & 1 \\
    1 & 0 & 1 & 1 & 1 \\
    0 & 1 & 0 & 1 & 1
  \end{pmatrix}.
\end{equation}

\subsubsection{\gate{CNOT} and \gate{SWAP} quantum gates and
  Gauss--Jordan elimination}
\label{sec:cnot-swap-quantum}

\gate{CNOT} and \gate{SWAP} quantum gates acts on elements of the
Pauli group by conjugation, $\hat P \to \hat U \hat P \hat U^\dagger$,
where $\hat P$ is an arbitrary element and
$\hat U \in \left\{\gate{CNOT},\ \gate{SWAP} \right\}$. Their action is
summarized in Table~\ref{tab:CNOT_SWAP_action}.
\begin{table}
  \centering
  \begin{tabularx}{0.5\textwidth}{@{}Xcc}
    \toprule
    Operation & Input & Output \\
    \midrule
    \multirow{4}*{\gate{CNOT}} & $\hat x_1$ & $\hat x_1 \hat x_2$ \\ \cmidrule{2-3}
              & $\hat x_2$ & $\hat x_2$ \\ \cmidrule{2-3}
              & $\hat z_1$ & $\hat z_1$ \\ \cmidrule{2-3}
              & $\hat z_2$ & $\hat z_1 \hat z_2$ \\
    \midrule
    \multirow{2}*{\gate{SWAP}} & $\hat x_1$ & $ \hat x_2$ \\ \cmidrule{2-3}
              & $\hat z_1$ & $\hat z_2$ \\
    \bottomrule
  \end{tabularx}
  \caption{Transformation properties of elements of the Pauli group
    under conjugation by \gate{CNOT} and \gate{SWAP}
    gates~\protect\cite[chap~10.5.2]{Nielsen:2010}. \gate{CNOT_{12}} has
    qubit 1 as the control and qubit 2 as the target, \gate{SWAP_{12}} acts
    on qubits 1 and 2.}
  \label{tab:CNOT_SWAP_action}
\end{table}

It must be noted that the \gate{SWAP} gate can be implemented as a
sequence of \gate{CNOT} gates; for our purposes, however, it is
convenient to define \gate{SWAP} as an independent operation to make a
perfect connection with the elementary matrix transformations (see
below). We also note that both gates are unitary and self-inverse, for
example, $\gate{CNOT} = \gate{CNOT}^\dagger = \gate{CNOT}^{-1}$, so
that in Table~\ref{tab:CNOT_SWAP_action} inputs and outputs can be
exchanged.

It is clear from Table~\ref{tab:CNOT_SWAP_action} that conjugation of
all operators from $\{\hat X_m\}$ with \gate{CNOT_{12}} -- here we
explicitly show what are the control and target qubits of the CNOT
operation -- is equivalent to adding the first row of the matrix
$\mathbf{M}$ [\latin{cf.} Eq.~\eqref{eq:example_M}] to its second row
\latin{modulo} 2. Similarly, conjugation with \gate{SWAP_{12}} is a
transposition of rows 1 and 2. By using \gate{CNOT_{ij}} or
\gate{SWAP_{ij}} we extend these operations to arbitrary qubits $i$
and $j$ and the corresponding rows of $\mathbf{M}$. That is, we
established a connection between certain transformations of Pauli
words and elementary matrix operations that are used in the
Gauss--Jordan elimination procedure. The latter applied to a matrix
$\mathbf{M}$ allows one to bring it to the reduced row-echelon form.
Again, considering the example of the $4\times 5$ matrix $\mathbf{M}$,
Eq.~\eqref{eq:example_M}, we obtain:
\begin{equation}
  \label{eq:example_Mrref}
  \mathbf{M}_\text{rref} = 
  \begin{pmatrix}
    1 & 0 & 0 & 0 & 1 \\
    0 & 1 & 0 & 0 & 1 \\
    0 & 0 & 1 & 0 & 0 \\
    0 & 0 & 0 & 1 & 0
  \end{pmatrix}.
\end{equation}
We shall call columns of $\mathbf{M}_{\text{rref}}$ with the leading 1
[\latin{e.g.} columns \numrange{1}{4} in Eq.~\eqref{eq:example_Mrref}]
the \emph{primary} columns (vectors). These columns correspond to the
Pauli elementary operators $\hat x_i$ in the binary representation.

A matrix $\mathbf{R}$ that brings $\mathbf{M}$ to its reduced row
echelon form, $\mathbf{RM} = \mathbf{M}_\text{rref}$, can be found by
applying Gauss--Jordan elimination to an augmented matrix
$(\mathbf{M} | \mathbf{E})$, where $\mathbf{E}$ is the $n\times n$
identity matrix. After the full execution of Gauss--Jordan elimination
it becomes $(\mathbf{M}_\text{rref} | \mathbf{R})$.

\subsubsection{$(2n-1)$ anti-commuting operators from a modified
  \gls{JW} transformation}
\label{sec:2n-1-anti}

By means of Gauss--Jordan elimination we mapped certain Pauli X~words
from $\{\hat X_m\}$ to the Pauli elementary operators $\hat x_i$. We
are now in a position to demonstrate how to convert them into a set
anti-commuting Pauli words with the odd number of $\hat y$ co-factors.
To this end let us recall the standard \acrlong{JW} construction,
which maps qubit (spin) operators $\hat x_i$ and $\hat y_i$ into
anti-commuting Majorana operators~\footnote{Majorana fermion operators
  $\hat C_i$ and $\hat D_i$, 
  are related to the ordinary (Dirac) fermion creation and
  annihilation operators as
  \begin{align*}
    {\hat a}_i & = \frac{1}{2}\left(\hat C_i + \I \hat D_i\right), \\
    {\hat a}_i^\dagger & = \frac{1}{2}\left(\hat C_i - \I \hat D_i\right).
  \end{align*}
}
\begin{align}
  \label{eq:JW_x_part}
  \hat x_i & \to  \hat C_i = \hat x_i \prod_{j=0}^{i-1} \hat z_j,\\
  \label{eq:JW_y_part}
  \hat y_i & \to  \hat D_i = \hat y_i \prod_{j=0}^{i-1} \hat z_j =
             \I\hat x_i \prod_{j=0}^i \hat z_j, 
\end{align}
where $0 \le i \le (n-1)$. There are $2n$ fully anti-commutative
operators $\{\hat C_i\}_{i=0}^{n-1}$ and $\{\hat D_i\}_{i=0}^{n-1}$.
The $(2n+1)$-th anti-commuting operator $\hat \Xi$ is a product of all
$\hat C_i$ and $\hat D_i$; it is equal to
\begin{equation}
  \label{eq:JW_last}
  \hat \Xi = \prod_{j=0}^{n-1} \hat z_j.
\end{equation}

Operators $\{\hat D_i\}_{i=0}^{n-1}$ already have the odd number of
$\hat y$ co-factors. The remaining $(n+1)$ operators have no $\hat y$
at all. However, it is possible to define $(n-1)$ operators with an
odd number of $\hat y$ as
\begin{equation}
  \label{eq:JW_modified}
  \hat F_i = \I \hat C_0 \hat C_i \hat \Xi =  \I \hat x_0 \hat x_i
  \prod_{j=i}^{n-1} \hat z_j = \hat x_0 \hat y_i
  \prod_{j=i+1}^{n-1} \hat z_j, \quad 1 \le i \le (n-1)   
\end{equation}
Direct calculations using the full anti-commutativity of $\hat C_i$,
$\hat D_i$, and $\hat \Xi$ show that $[\hat F_i, \hat F_j]_{+} = 0$
for $i \ne j$ and $[\hat D_i, \hat F_j]_{+} = 0$. Thus, we have
constructed a set of $(2n-1)$ fully anti-commutative operators
$\{\hat D_i\}_{i=0}^{(n-1)}$ and $\{\hat F_i\}_{i=1}^{(n-1)}$ which
all have the odd number of $\hat y$ co-factors.

\subsubsection{A set of anti-commuting operators from $\{\hat X_m\}$}
\label{sec:set-anti-commuting}

Eqs.~\eqref{eq:JW_y_part} and \eqref{eq:JW_modified} demonstrate how
to convert Pauli words $\{\hat x_i\}_{j=0}^{(n-1)}$ and
$\{\hat x_0\hat x_i\}_{j=1}^{(n-1)}$ into a $(2n-1)$-member set of
fully anti-commutative operators by appending Z~words
$\left\{\prod_{j=0}^i\hat z_j\right\}_{i=0}^{n-1}$ and
$\left\{\prod_{j=i}^{n-1}\hat z_j\right\}_{i=1}^{n-1}$, respectively.
These operators are defined in the representation in which the matrix
$\mathbf{M}$ is brought to the reduced row-echelon form. The primary
columns of the matrix $\mathbf{M}_\text{rref}$ correspond to a subset
of $\{\hat x_i\}_{j=0}^{(n-1)}$. In turn, columns of
$\mathbf{M}_\text{rref}$ that have the binary representation
$(\mathbf{1}\, 0\, \dots 0\, 1_i\, 0\, \dots\, 0)^T$ are associated
with a subset of $\{\hat x_0\hat x_i\}_{j=1}^{(n-1)}$; we shall refer
to such vectors as ``secondary''. In the example~\eqref{eq:example_Mrref} the
last column represents a single secondary vector.

Primary and secondary Z operators (and corresponding vectors) are
defined as partners of primary/secondary X operators in
Eqs.~\eqref{eq:JW_y_part} and \eqref{eq:JW_modified}. We collect all
definitions in Table~\ref{tab:XZ_anticomm}.
\begin{table}
  \centering
  \caption{Primary and secondary X and Z operators whose products
    $\hat T_k = \hat X_k \odot \hat Z_k$ are fully anti-commutative.
    The total length of the strings in the binary representation is
    $n$, which is the number of qubits.}
  \begin{tabularx}{1.0\linewidth}{@{}Xccc}
    \toprule
    Type & Classification &  Operator form & Binary representation \\
    \midrule
    \multirow{2}*{X~words} & Primary   & $\hat x_i$ & $(0\, \dots\, 0\, 1_i\, 0\, \dots\, 0)^T$ \\ \cmidrule{2-4}
         & Secondary & $\hat x_0\hat x_i$ & $(1\, 0\, \dots\, 0\, 1_i\, 0\, \dots\, 0)^T$ \\
    \midrule
    \multirow{2}*{Z~words} & Primary   & $\prod_{j=0}^i\hat z_j$ & $(1\, \dots\, 1_i\, 0\, \dots\, 0)^T$ \\ \cmidrule{2-4}
         & Secondary   & $\prod_{j=i}^{(n-1)}\hat z_j$ & $(0\, \dots\, 0\, 1_i\, \dots\, 1)^T$ \\
    \bottomrule
  \end{tabularx}
  \label{tab:XZ_anticomm}
\end{table}

It is important to emphasize that \emph{only primary and secondary
  vectors of $\mathbf{M}_\text{rref}$ can be used to create the
  anti-commutative system}. The number of such vectors is uniquely
determined by $\mathbf{M}$ and, in turn, by the composition and
ordering of $\{\hat X_m\}$.


Our final task is to return to the original basis. For X~words this
operation is trivial: the inverse transform $\mathbf{R}^{-1}$ brings
the primary and secondary columns of $\mathbf{M}_\text{rref}$ back to
the corresponding columns of $\mathbf{M}$. These columns hence
determine a subset $\{\hat X_{m_l}\}$ of Pauli words in $\{\hat X_m\}$
that participate in formation of the anti-commutative system. The
solution is slightly more involved for Z~words. A careful inspection
of Table~\ref{tab:CNOT_SWAP_action} shows that the \gate{CNOT}
operation acts on Z~words as a \emph{transposed} row-addition
elementary operation. We have to also reverse the ordering of
operations in the Gauss--Jordan elimination, so that if
$\mathbf{R} = \mathbf{R}_1 \cdots \mathbf{R}_k$ then the desired
transformation is $(\mathbf{R}_k^{-1})^T \cdots (\mathbf{R}_1^{-1})^T$
and since $\mathbf{R}_j^{-1} = \mathbf{R}_j$ (modulo 2), it is
equivalent to
$\mathbf{R}_k^T \cdots \mathbf{R}_1^T = \mathbf{R}^T$. Thus, one
has to apply $\mathbf{R}^T$ to the primary and secondary binary
Z~vectors (see rows \numrange{4}{5} of Table~\ref{tab:XZ_anticomm}) to
generate the corresponding Z vectors in the original representation
and, subsequently, to recover Z~words that are partners of X~words in
$\{\hat X_{m_l}\}$. The resulting operators
$\hat T_{m_l} = \hat X_{m_l} \odot \hat Z_{m_l}$ are all mutually
anti-commutative, $[\hat T_{k'}, \hat T_k]_{+} = 0$ for $k' \ne k$,
because anti-commutativity is preserved by conjugation with
\gate{CNOT} and \gate{SWAP} gates. Completing the example given by
Eqs.~\eqref{eq:example_X_set}, \eqref{eq:example_M}, and
\eqref{eq:example_Mrref}, we find the following anti-commuting system:
\begin{equation}
  \label{eq:example_anticomm_set}
  \{ {\hat y}_{0}{\hat z}_{1}{\hat x}_{2}{\hat z}_{3}, \
  {\hat y}_{1}{\hat z}_{2}{\hat x}_{3}, \
  {\hat x}_{0}{\hat x}_{1}{\hat y}_{2}{\hat z}_{3},\ 
  {\hat z}_{0}{\hat x}_{1}{\hat x}_{2}{\hat y}_{3},\ 
  {\hat x}_{0}{\hat y}_{1}{\hat x}_{2}{\hat x}_{3} \}.
\end{equation}

Every $\hat T_{m_l}$ contains the odd number of $\hat y$ factors,
albeit not necessarily a \emph{single} such factor like operators in
the standard system $\{\hat D_i\} \cup \{\hat F_i\}$ [see
Eqs.~\eqref{eq:JW_y_part} and \eqref{eq:JW_modified}]. This remarkable
fact can be proven as follows. Pauli words with the odd number of
$\hat y$ factors are \textit{imaginary} operators -- their matrices
have purely imaginary elements in the tensor-product basis of
eigenstates of $\hat z_i$, $0 \le i \le (n-1)$. \gate{CNOT} and
\gate{SWAP} gates are purely real matrices in the same basis, so that
the imaginary matrices (hence, operators) in the standard basis
remains imaginary under conjugation with any of these operators.

\subsubsection{Practical considerations}
\label{sec:pract-cons}

There is still a great deal of flexibility in the algorithm. The
resulting set of anti-commuting Paulis depends on the ordering of
columns of the matrix $\mathbf{M}$ and hence, the ordering of
operators in $\{\hat X_m\}$. We always take an operator with the
largest gradient (or another ``importance measure'', see
Ref.~\citenum{Ryabinkin:2021/qst/024012}) as first, which ultimately
guarantees the convergence of the \gls{iQCC}
procedure~\cite{Ryabinkin:2020/jctc/1055}. The whole $\{\hat X_m\}$,
however, may not contain enough operators to find $n$ primary and
$(n-1)$ secondary vectors to construct the maximal anti-commutative
set. In such a situation we do \emph{not} construct complementary
operators as they will have zero gradients and their energy impact is
unclear. On the other hand, in the case of extremely large
$\{\hat X_m\}$ it may be necessary to drop some operators from
consideration, and those with the smallest gradients are the first
candidates. Operators from $\{\hat X_m\}$ may have vanishing gradients
because of special symmetry of the qubit reference vector
$\ket{0}$---for example, when $\ket{0}$ is an eigenvector of
$\hat I_k(\mathbf{z})$ for some $k >0$ with 0 as an
eigenvalue\footnote{The corresponding gradient value $\omega_k = 0$
  according to Eq.~\eqref{eq:omega_k}}. In what follows we do not drop
operators from $\{\hat X_m\}$ with vanishing gradients to simplify the
discussion.

\subsubsection{Algorithm summary}
\label{sec:summ-algor-flowch}


\begin{algorithm*}
  \DontPrintSemicolon
  \SetKwInOut{Input}{Input} %
  \SetKwInOut{Output}{Output} %
  
  \Input{A sequence of X~words $\{\hat X_m\}$} %
  \Output{$\{\hat X_{m_l}\} \subset \{\hat X_m\}$ and
    $\{\hat Z_{m_l}\}$ such that
    $\hat T_{m_l} = \hat X_{m_l} \odot \hat Z_{m_l}$ and
    $\left[\hat T_{l}, \hat T_{l'}\right]_{+} = 2\delta_{ll'}$} %
  \BlankLine %
  
  \ForEach(\tcp*[h]{form a $\mathbf{M}$ matrix}){${\hat X}_k \in
    \{\hat X_m\}$}{ %
    $\mathbf{M}[:, k] = \text{BinaryRepresentation}(\hat X_k)$ \; %
    $k \gets k+1$ \; %
  } %
  $\mathbf{M}_\text{rref},\, \mathbf{R} \gets \text{rref}(\mathbf{M})$
  \tcp*[h]{Gauss--Jordan elimination} \; %

  Identify primary and secondary columns of $\mathbf{M}_\text{rref}$
  \; %

  $\{\hat X_{m_l}\}$: select from $\{\hat X_m\}$ operators that
  correspond to primary and secondary column indices\; %

  Construct primary and secondary binary vectors $\{Z_k\}$ for every
  primary and secondary column in $\mathbf{M}_\text{rref}$ using
  Table~\ref{tab:XZ_anticomm}. \; %
  
  Transform $\{Z_k\}$ from the standard to the original basis:
  $Z_k \gets \mathbf{R}^T Z_k$\; %
  
  $\{\hat Z_{m_l}\}$: Convert binary vectors $\{Z_k\}$ back to the
  operator form \;
  
  \caption{A set of anti-commuting Pauli words from a set of X~words}
    \label{alg:1}
\end{algorithm*}

\subsection{Perturbative correction for QCC-ILCAP}
\label{sec:pert-corr-qcc}

The size of any anti-commuting system $\sim n$ is fundamentally
limited by the qubit dimensionality of a system. Since the number of
groups of operators with non-zero gradients in the qubit image of a
fermionic Hamiltonian~\eqref{eq:qubit_H} is $O(n^4)$, the bulk of
energy lowering due to correlation will \emph{not} be captured by the
QCC-\gls{ILCAP} Ansatz in a single iteration. Though a fully iterative
procedure is possible, a single application of QCC-\gls{ILCAP} as a
pre- and especially post-processing technique calls for a completeness
correction. As such, we propose a Brillouin--Wigner perturbation
theory at the second order motivated by theoretical simplicity and
computational efficiency of this approach. Namely, the unitary
QCC-\gls{ILCAP} Ansatz is equivalent to a linear parametrization of a
variational wave function, as follows from Eq.~\eqref{eq:U_ilcap}.
Thus, the energy functional~\eqref{eq:QCC_energy_expr} is a quadratic
form
\begin{equation}
  \label{eq:energy_ilcap}
  E_\text{ILCAP}(\mathbf{C}) = \mathbf{C}^\dagger \mathbf{H}
  \mathbf{C}, 
\end{equation}
where $\mathbf{H}$ is an $(M+1) \times (M+1)$ matrix with elements
\begin{equation}
  \label{eq:H_matrix}
  H_{k'k} = \left\{
    \begin{array}{cc}
      \braket{0|\hat H|0}, & k' = k = 0, \\
      \I\braket{0|\hat T_{k'}\hat H |0}, & 1 \le k' \le M,\, k = 0, \\
      -\I\braket{0|\hat H \hat T_k|0}, & 1 \le k \le M,\, k' = 0, \\
      \braket{0|\hat T_{k'} \hat H \hat T_k|0}, & 1 \le k',k \le M,
    \end{array}
  \right.
\end{equation}
$\hat T_{k'}$ and $\hat T_{k}$ are operators from \gls{ILCAP} and
components of a vector $\mathbf{C}$ are related to cluster amplitudes
$(t, \boldsymbol\alpha)$ as
\begin{align}
  \label{eq:C0_eq}
  C_0 & = \cos\left(\frac{t}{2}\right), \\
  \label{eq:Ci_eq}
  C_i & = \alpha_i\sin\left(\frac{t}{2}\right),\, 1 \le i \le M.
\end{align}
Choosing anti-commuting generators $\hat T_k$ with the odd number of
$\hat y$ ensures the matrix elements~\eqref{eq:H_matrix} are all real.
Extrema of the functional~\eqref{eq:energy_ilcap} are eigenvectors of
$\mathbf{H}$ that can be chosen orthonormal; the one which corresponds
to the lowest eigenvalue is used to recover amplitudes by
Eqs.~\eqref{eq:C0_eq} and \eqref{eq:Ci_eq}.

The relation of the energy minimization with the QCC-\gls{ILCAP}
Ansatz with the eigenvalue problem
$\mathbf{H}\mathbf{C} = E\mathbf{C}$ provides a unique opportunity to
apply a Brillouin--Wigner perturbation theory \latin{via} the L\"owdin
partitioning~\cite{Lowdin:1964/jms/326}. Consider a rectangular matrix
$\mathbf{b}$ with elements
\begin{equation}
  \label{eq:b_matrix}
  b_{km} = \left\{
    \begin{array}{lc}
      \braket{0|\hat H \hat X_m|0}, & k = 0, \\
      \I\braket{0|\hat T_{k} \hat H \hat X_m |0}, & 1 \le k \le M,
    \end{array}
  \right.
\end{equation}
and a diagonal matrix $\mathbf{D}$ with elements
$\braket{0|\hat X_m \hat H \hat X_m|0}$, where $\hat X_m$ are
operators from the Ising decomposition of $\hat H$ that were
\emph{not} used to construct \gls{ILCAP}. Then the Brillouin--Wigner
ground-state energy estimate is the lowest eigenvalue of an
energy-dependent matrix
\begin{equation}
  \label{eq:E_BW}
  \mathbf{H}^\text{eff}(E) = \mathbf{H} - \mathbf{b} (\mathbf{D} -
  E)^{-1} \mathbf{b}^\dagger. 
\end{equation}
Due to the energy dependence of $\mathbf{H}^\text{eff}$, iterations
are necessary to obtain the ground-state energy estimate,
$E_\text{ILCAP+BW}$. However, computational time and memory storage
requirements for both $\mathbf{b}$ and $\mathbf{D}$ matrices are
linearly proportional to the size of a Hamiltonian (more accurately,
$\{\hat X_m\}$ set).

To perform the completeness correction one can alternatively apply the
Epstein--Nesbet perturbation theory~\cite{Ryabinkin:2021/qst/024012}
after the dressing of the current Hamiltonian (see
Sec.~\ref{sec:outl-glsiqcc-meth}) with the optimized QCC-\gls{ILCAP}
transformation. While this approach is not viable as a post-processing
technique, it cross-validates applicability of the perturbation
theory.


\section{Results and discussion}
\label{sec:results-discussions}

\subsection{General setup}
\label{sec:general-setup}

To illustrate our developments we compute potential energy curves for
the symmetric stretch of an \ce{H2O} molecule and dissociation of
\ce{N2}. Both processes describe a graduate transition from weak to
strong correlation; a robust method must not break down upon this
transition. Two considered models differ in their qubit
dimensionality: the first one is a mid-scale 36-qubit problem while
the second is a large-scale 56 qubit one.

For each molecule we consider the following schemes:
\begin{enumerate}
\item {\bfseries \gls{QCC}-\gls{ILCAP} as a pre-processing technique}.
  We optimize the \gls{QCC}-\gls{ILCAP} Ansatz~\eqref{eq:U_ilcap} to
  find an approximate ground state
  $\ket{\Psi}_\text{ILCAP} = \hat U_\text{ILCAP}(t_\text{opt},
  \boldsymbol\alpha_\text{opt}) \ket{0}$, the corresponding
  ground-state energy estimate
  $E_\text{ILCAP} = \braket{\Psi_\text{ILCAP} | \hat H
    |\Psi_\text{ILCAP}}$, and its Brillouin--Wigner-corrected
  counterpart, $E_\text{ILCAP+BW}$, computed as described in
  Sec.~\ref{sec:pert-corr-qcc}. Here $\hat H$ is a fermionic
  Hamiltonian~\eqref{eq:qe_ham} converted to the qubit representation.
  At this point, the \gls{QCC}-\gls{ILCAP} approach can be considered
  as a quantum-inspired method for adding the electron correlation to
  a mean-field description (encoded as $\ket{0}$) of a molecular
  ground state. Next, aiming at the use of the \gls{QCC}-\gls{ILCAP}
  optimized unitary to facilitate electronic structure calculations on
  \gls{NISQ} devices, we define the \gls{ILCAP}-dressed Hamiltonian,
  $\hat H_\text{ILCAP} = \hat U^\dagger_\text{ILCAP}(t_\text{opt},
  \boldsymbol\alpha_\text{opt}) \hat H \hat
  U_\text{ILCAP}(t_\text{opt}, \boldsymbol\alpha_\text{opt})$. To
  simulate the further use of $\hat H_\text{ILCAP}$ on a quantum
  computer, we evaluate the Epstein--Nesbet perturbation theory
  correction (denoted as $E_\text{ILCAP+EN}$) to the ground-state
  energy estimate
  $\braket{0|\hat H_\text{ILCAP}|0} \equiv E_\text{ILCAP}$ as
  described in Ref.~\citenum{Ryabinkin:2021/qst/024012}.

\item {\bfseries \gls{QCC}-\gls{ILCAP} as a post-processing
    technique}. We perform several rounds of the \gls{iQCC} procedure
  as described in Sec.~\ref{sec:outl-glsiqcc-meth} and take the last
  dressed Hamiltonian, $\hat H^{(i+1)}$ ($i \ge 1$) as the starting
  one for the application of the \gls{QCC}-\gls{ILCAP} Ansatz. After
  optimization of amplitudes $(t, \boldsymbol\alpha)$ we consider the
  final energy, $E_\text{QCC(i)+ILCAP}$ as well as its
  Brillouin--Wigner-corrected partner, $E_\text{QCC(i)+ILCAP+BW}$ as
  new ground-state estimates and compare them to the ``bare''
  \gls{iQCC} energies and their Epstein--Nesbet-corrected counterparts
  denoted as $E_\text{QCC(i)}$ and $E_\text{QCC(i)+EN}$, respectively.
\end{enumerate}

For both molecules we prepared a set of \gls{RHF} \glspl{MO} which
were subsequently used to compute one- and two-electron integrals in
the active space.
These calculations were carried out using a modified version of
\textsc{gamess}~\cite{gamessus, gamessus-2}, from Sep 30, 2019 (R2).

The common setup for the \gls{iQCC} method is as follows. Generators
were ranked according to the ``optimal amplitude'' defined by Eq.~33
of Ref.~\citenum{Ryabinkin:2021/qst/024012} as it works slightly
better than the gradient ranking in the case of strong correlation
(highly stretched geometries). To stay on a singlet solution along the
whole potential curve we added a penalty operator $\hat W$ to the
initial Hamiltonians at every nuclear configuration,
\begin{align}
  \label{eq:H_plus_penalty}
  \hat H & \to \hat H + \frac{\mu}{2} \hat W, \\
  \label{eq:penalty_op}
  \hat W & = \hat S^2 - (2s+1)\hat S_z + s^2,
\end{align}
where $\hat S^2$ and $\hat S_z$ are the qubit images of the total spin
squared operator and its $z$ projection, respectively, $s = 0$ is a
spin quantum number for a singlet state, and $\mu$ is a penalty
parameter.

\subsection{\ce{H2O} simulations}
\label{sec:ceh2o-simulations}

The \Gls{RHF} \glspl{MO} were expanded in 6-31G(d) atomic basis
set~\cite{Hehre:1972/jcp/2257} with a six-component (Cartesian) $d$
polarisation function assuming $C_{2v}$ symmetry at selected \ce{O-H}
distances and fixed $\angle \ce{HOH} = \SI{107.60}{\degree}$. The
lowest-energy \gls{MO} that correlates with an $1s$ atomic orbital of
oxygen atom was frozen while the remaining 18 orbitals were considered
as active. This leads to a 36-qubit Hamiltonian~\eqref{eq:qubit_H}
with \num{41915} qubit terms; all terms with coefficients smaller than
\num{e-8} in magnitude were discarded. The spin-penalty strength
parameter $\mu$ [see Eq.~\eqref{eq:H_plus_penalty}] was set to
\num{0.025}; the number of terms in the penalized operator is
\num{42527}. The qubit reference vector~\eqref{eq:qubit_ref}
represents an eight-electron closed-shell singlet state ($n_e = 8$).

\begin{figure*}
  \centering %
  \includegraphics[width=1.0\textwidth]{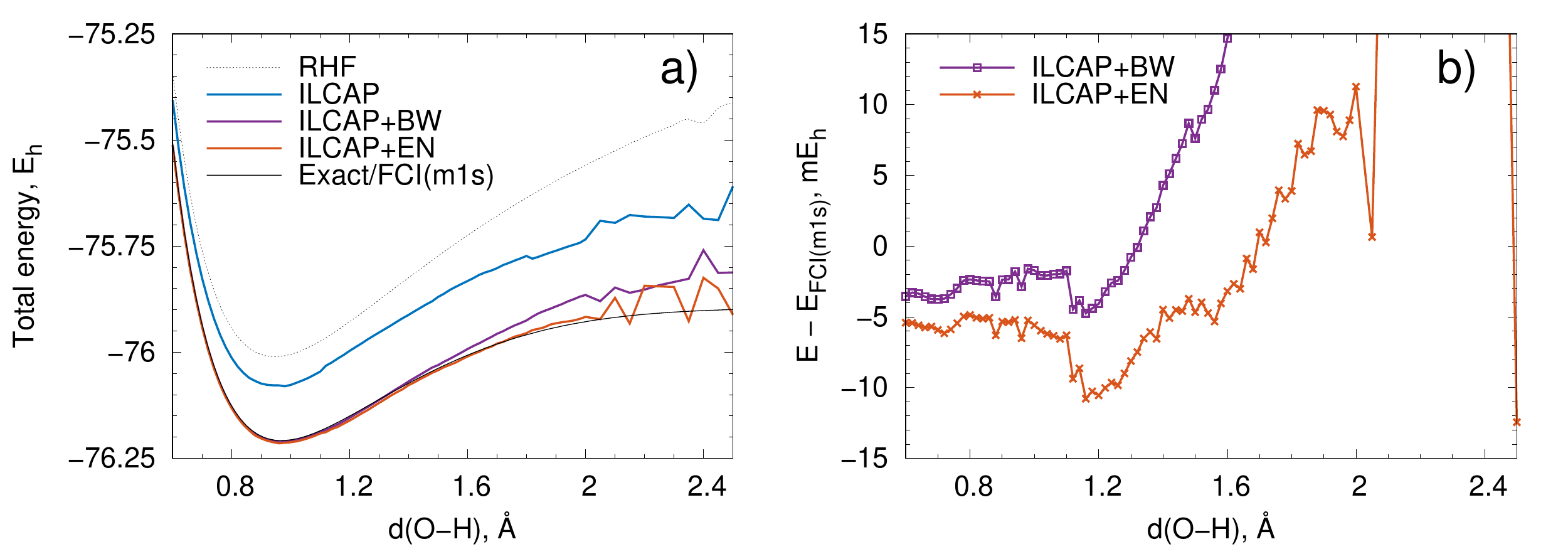}
  \caption{a) Potential energy curves for symmetric \ce{O-H}
    stretching of \ce{H2O} molecule. b) Deviations of the energy (in
    \si{\milli\hartree}) from the exact-diagonalization result denoted
    as FCI(m1s)}
  \label{fig:h2o_symm_pec}
\end{figure*}

\paragraph{QCC-ILCAP as a pre-processing technique.} For every value
of $d(\ce{O-H})$ we report three ground-state energy
estimates\footnote{The nuclear-nuclear repulsion energy $V_{nn}$ is
  added for the total electronic energy at every nuclear
  configuration.}, $E_\text{ILCAP}$, $E_\text{ILCAP+BW}$, and
$E_\text{ILCAP+EN}$ computed as described in
Sec.~\ref{sec:general-setup}. We compare them against \gls{CASCI}
values (denoted as $E_\text{FCI(m1s)}$) computed by the Davidson
diagonalization of the fermionic Hamiltonian in the determinant basis
in \textsc{gamess}. In $C_{2v}$ symmetry this basis contains
\num{2342224} determinants with $s_z = 0$. The resulting potential
energy curves are shown in Fig.~\ref{fig:h2o_symm_pec}.

As follows from Fig.~\ref{fig:h2o_symm_pec}a, \gls{QCC}-\gls{ILCAP}
provides a reasonably smooth potential energy curve for
$d(\ce{O-H}) \lessapprox \SI{1.8}{\angstrom}$, while for ILCAP+BW and
ILCAP+EN schemes this range extends to
$d(\ce{O-H}) \approx \SI{2.0}{\angstrom}$. However, closer inspection
of seemingly good curves for those two schemes in a range
\SIrange{0.6}{2.0}{\angstrom} (see Fig.~\ref{fig:h2o_symm_pec}b)
reveals a saw-tooth pattern with oscillations of a few millihartee in
magnitude superimposed on gradually increasing deviation from the
exact result. Such non-smooth potential energy curves for the
\gls{QCC} method were already reported~\cite{Lang:2021/jctc/66} and
were attributed to re-ordering of generators employed in the
construction of the \gls{QCC} Ansatz. Here we construct a
\gls{QCC}-\gls{ILCAP} of the maximal possible size using \emph{all}
generators with \latin{a priori} non-zero gradients; despite this,
``kinks'' still have an appreciable magnitude greater than the
so-called ``chemical accuracy'' of $\sim \SI{1}{\milli\hartree}$.
Kinks of smaller amplitude are likely due to re-ordering of generators
with non-maximal importance measure. However, the first large kink
near $d(\ce{O-H}) \approx \SI{2.1}{\angstrom}$ is due to the change of
the leading (top-ranked, first) generator. The top-ranked generator
(out of 47 included in \gls{ILCAP}) at
$d(\ce{O-H}) = \SI{2.05}{\angstrom}$ has
$\hat x_6 \hat x_7 \hat x_{10} \hat x_{11}$ parental X~word, while at
$d(\ce{O-H}) = \SI{2.1}{\angstrom}$ the top-ranked generator (out of
43 in the whole \gls{ILCAP}) stems from the
$\hat x_6 \hat x_7 \hat x_8 \hat x_9$ word which again becomes
$\hat x_6 \hat x_7 \hat x_{10} \hat x_{11}$ at
$d(\ce{O-H}) = \SI{2.15}{\angstrom}$ with 50 operators in the entire
\gls{ILCAP}. These substantial changes in composition of \gls{ILCAP}
are reflected in large variations of $E_\text{ILCAP}$,
$E_\text{ILCAP+BW}$, and $E_\text{ILCAP+EN}$ in the small range
\SIrange{2.05}{2.15}{\angstrom} of \ce{O-H} distances.

Overall, the \gls{QCC}-\gls{ILCAP} method captures less than a half of
the total correlation energy and requires the perturbation correction
to account for the rest. The Brillouin--Wigner approach indeed brings
the total energy to be a few millihartree away from the exact value,
but only if the electron correlation is not too strong. The ILCAP+EN
scheme performs slightly better but still suffers from large
variations near the dissociation limit.

\begin{figure*}
  \centering %
  \includegraphics[width=1.0\textwidth]{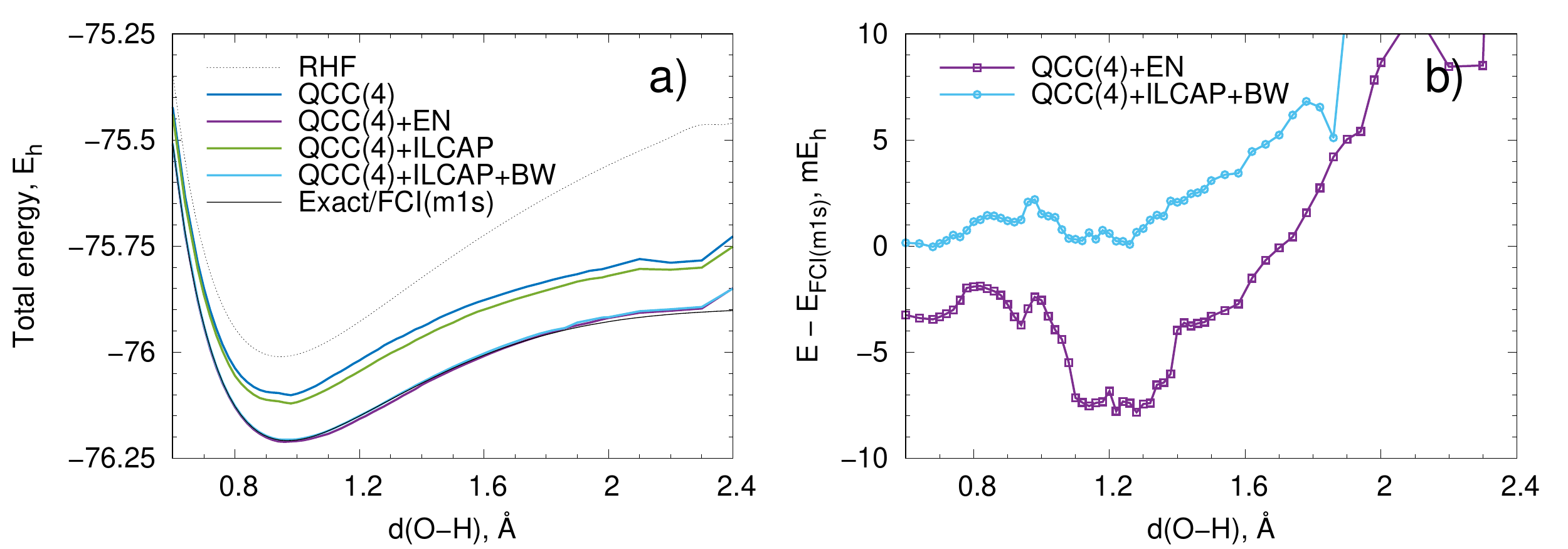}
  \caption{Same as for Fig.~\ref{fig:h2o_symm_pec}.}
  \label{fig:h2o_symm_pec_pp}
\end{figure*}

\paragraph{QCC-ILCAP as a post-processing technique.} We have
performed four ($i=4$) \gls{iQCC} iterations with $L = 12$ generators
in Eq.~\eqref{eq:U_QCC} and constructed the \gls{ILCAP} Ansatz based
on the Ising decomposition of $\hat H^{(5)}$ Hamiltonian [see
Eq.~\eqref{eq:dressing}]. We report four energy estimates,
$E_\text{QCC(4)} = \braket{0|\hat H^{(5)}|0}$, $E_\text{QCC(4)+EN}$,
$E_\text{QCC(4)+ILCAP}$, and $E_\text{QCC(4)+ILCAP+BW}$, which are
shown in Fig.~\ref{fig:h2o_symm_pec_pp}.
Figure~\ref{fig:h2o_symm_pec_pp}a demonstrates that preliminary
\gls{iQCC} iterations account for \SIrange{40}{60}{\percent} of the
correlation energy while the subsequent \gls{ILCAP} treatment adds
relatively small amount of the remaining correlation energy. Both
perturbation corrections, the Epstein-Nesbet one on top of QCC(4) and
Brillouin--Wigner on top of the QCC(4)+ILCAP wave function provide
comparable accuracy. The QCC(4)+ILCAP+BW scheme appears to be superior
to the QCC(4)+EN counterpart giving a smoother curve with a deviation
of \SIrange{0}{5}{\milli\hartree} in the range of $d(\ce{O-H})$
\SIrange{0.6}{1.8}{\angstrom}. Comparison of
Figs.~\ref{fig:h2o_symm_pec} and \ref{fig:h2o_symm_pec_pp} shows that
the preceding \gls{iQCC} iterations not only mitigate the problem of
non-smooth potential energy curves but also help ILCAP and especially
ILCAP+BW schemes to provide more accurate energies in a wider range of
\ce{O-H} distances.

\subsection{\ce{N_2} simulations}
\label{sec:cN2-simulations}

The \gls{RHF} orbitals were expanded in Dunning's cc-pVDZ atomic basis
set~\cite{Dunning:1989/jcp/1007} assuming $D_{2h}$ symmetry of a
molecule. All 28 orbitals and 14 electrons were taken as active, so
the second-quantized Hamiltonians~\eqref{eq:qe_ham} represented a
\gls{FCI} problem; the corresponding 56-qubit Hamiltonians contained
\num{107881} Pauli terms at every value of $d(\ce{N-N})$. Terms
smaller than \num{e-7} were dropped from the initial and dressed
Hamiltonians. The spin-penalty strength $\mu$ [see
Eqs.~\eqref{eq:H_plus_penalty} and \eqref{eq:penalty_op}] was set to
\num{0.125}, the penalized Hamiltonian had \num{109393} terms.

The \gls{FCI} energies were taken from
Ref.~\citenum{Wang:2019/jctc/3558} as the conventional \gls{FCI}
calculations are not possible even with supercomputer resources. The
\gls{FCI} energy estimates are believed to be accurate to
\SI{1e-6}{\hartree}.

\begin{figure*}
  \centering %
  \includegraphics[width=1.0\textwidth]{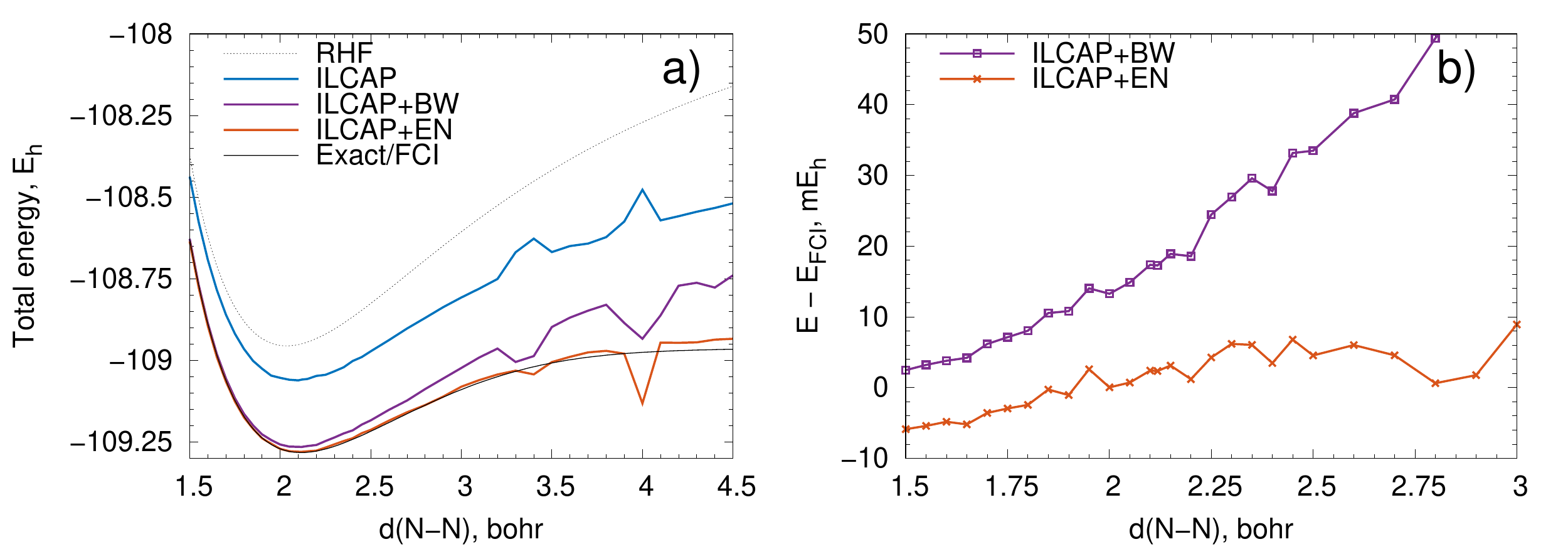}
  \caption{Same as for Fig.~\ref{fig:h2o_symm_pec} but for the \ce{N2}
    molecule stretching.}
  \label{fig:N2_stretch_pre}
\end{figure*}

\paragraph{QCC-ILCAP as a pre-processing technique.} $E_\text{ILCAP}$,
$E_\text{ILCAP+BW}$, and $E_\text{ILCAP+EN}$ along with \gls{FCI} and
\gls{RHF} curves are plotted in Fig.~\ref{fig:N2_stretch_pre}.
Compared with Fig.~\ref{fig:h2o_symm_pec} kinks on \gls{ILCAP},
\gls{ILCAP}+BW and \gls{ILCAP}+EN curves appear earlier, already at
$\sim 1.5 R_e$, and have larger amplitudes. This is expected as
correlation is stronger for a triple-bond breaking process. Just like
in the case of \ce{H2O} the most noticeable kinks are related to
changes in the leading (top-ranked) generator used to construct the
\gls{QCC}-\gls{ILCAP} Ansatz: at $d(\ce{N-N}) = \SI{3.2}{\bohr}$ the
top-ranked generator stems from the
${\hat x}_{10}{\hat x}_{11}{\hat x}_{14}{\hat x}_{15}$ X~word, which
changes into ${\hat x}_{10}{\hat x}_{13}{\hat x}_{15}{\hat x}_{16}$ at
$d(\ce{N-N}) = \SI{3.3}{\bohr}$, then to
${\hat x}_{12}{\hat x}_{13}{\hat x}_{14}{\hat x}_{15}$ at
$d(\ce{N-N}) = \SI{3.4}{\bohr}$ and finally returns back to
${\hat x}_{10}{\hat x}_{11}{\hat x}_{14}{\hat x}_{15}$ at
$d(\ce{N-N}) = \SI{3.5}{\bohr}$.

Quantitatively, the deviations from the \gls{FCI} values are
\numrange{2}{5} times larger that those in
Fig.~\ref{fig:h2o_symm_pec}; however, they are accumulated more
uniformly. Somewhat surprisingly, the \gls{ILCAP}+EN scheme based on
the ILCAP-dressed Hamiltonian is closer to the \gls{FCI} reference
than any other schemes, staying within \SIrange[range-phrase={ to
}]{-6}{6}{\milli\hartree} in the range of \ce{N-N} distances
\SIrange{1.5}{3.0}{\bohr}. Perhaps, this is because the second-order
perturbation theory based on the dressed Hamiltonian contains
higher-order contributions in terms of the original fermionic
Hamiltonian which are more important for triple-bond breaking process
than for the simultaneous breaking of only two bonds in \ce{H2O}.

\paragraph{QCC-ILCAP as a post-processing technique.} We performed
$i=4$ \gls{iQCC} iterations with $L = 14$ generators, and the
\gls{ILCAP} Ansatz was constructed using the $\hat H^{(5)}$ dressed
Hamiltonian. The number of terms in the Ising
decomposition~\eqref{eq:Ising_decomp} of $\hat H^{(5)}$ varied with
$d(\ce{N-N})$ peaking at \num{45e6}.

\begin{figure*}
  \centering %
  \includegraphics[width=1.0\textwidth]{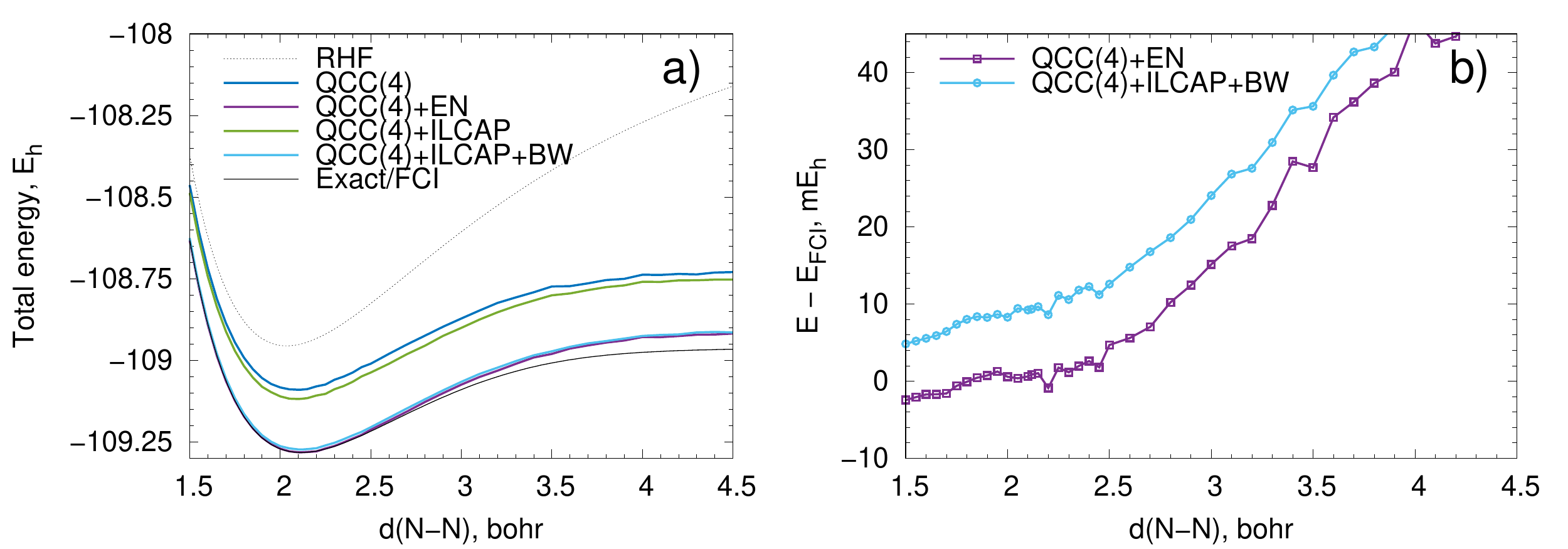}
  \caption{Same as for Fig.~\ref{fig:h2o_symm_pec} but for the \ce{N2}
    molecule stretching.}
  \label{fig:N2_stretch_pp}
\end{figure*}

Potential energy curves for four \gls{QCC}-based schemes,
$E_\text{QCC(4)} = \braket{0|\hat H^{(5)}|0}$, $E_\text{QCC(4)+EN}$,
$E_\text{QCC(4)+ILCAP}$, and $E_\text{QCC(4)+ILCAP+BW}$ are displayed
in Fig.~\ref{fig:N2_stretch_pp}.
Preceding \gls{iQCC} iterations solved the problem of large kinks on
the potential energy curves for all considered schemes. As evident
from Fig.~\ref{fig:N2_stretch_pp}b, the remaining variations are of
the order of \SI{1}{\milli\hartree} or less, with increasing deviation
from \gls{FCI} upon increasing of the \ce{N-N} distance. The overall
quality of the two best variants, $E_\text{QCC(4)+ILCAP+BW}$ or
$E_\text{QCC(4)+EN}$, is very close; the former is shifted upwards
compared to the \gls{FCI} reference. In order to assess the quality of
\emph{relative} energies, we fitted the total energies in the range of
\ce{N-N} distances \SIrange{1.5}{3.0}{\bohr} to the Morse potential
curve,
\begin{equation}
  \label{eq:Morse_fit}
  E(r) = D_e\left(1 - \E^{-a(r-r_e)}\right)^2 + E_\text{min},
\end{equation}
where $D_e$ is the dissociation energy, $r_e$ is the position of the
minimum, $E_\text{min} = E(r_e)$, and $a$ is a parameter related to
the force constant. From this fit we computed the spectroscopic
properties, namely, the harmonic frequency $\omega_e$ and the first
anharmonicity constant $\omega_ex_e$ as
\begin{align}
  \label{eq:w_e}
  \omega_e & = a\sqrt{\frac{2D_e}{\mu}} \\
  \omega_ex_e & = \frac{\omega_e^2}{4D_e},
\end{align}
where $\mu$ is the reduced mass of a \ce{^{14}N2} species,
$\mu = m_{\ce{N}}/2 = \SI{7.00155}{\amu}$. All quantities are
collected in Table~\ref{tab:N2_spectroscopic_params}.
\begin{table*}
  \centering
  \caption{Spectroscopic constants of the \ce{N2} from the Morse
    fitting, Eq.~\eqref{eq:Morse_fit}.}
  \begin{minipage}{1.0\linewidth}
    \sisetup{ %
      table-auto-round, %
      round-mode = places} %
    \begin{tabularx}{1.0\linewidth}{@{}XS[table-format=+6.4]S[table-format=1.4]S[table-format=1.3]S[table-format=1.3]S[table-format=1.2e-1]S[table-format=4.0]S[table-format=2.1]@{}}
      \toprule %
      Method          & {$E_\text{min}$} & {$D_e$}         & {$r_e$, \si{\bohr}}  & {$a$, \si{\bohr^{-1}}} & {RMS of residuals} & {$\omega_e$}  & {$\omega_ex_e$}            \\
      \cmidrule{2-3}                                                                                           \cmidrule{7-8}
                      & \multicolumn{2}{c}{\si{\hartree}} &                      &                        &                    & \multicolumn{2}{c}{\si{\centi\meter^{-1}}} \\
      \midrule %
      {\small QCC(4)+EN}       & -109.2819    & 0.4434          & 2.112                   & 1.295                  & 9.67524e-4        & 2369           & 14.4 \\
      {\small QCC(4)+ILCAP+BW} & -109.2732    & 0.4402          & 2.110                   & 1.301                  & 7.6511e-4          & 2371           & 14.6 \\
      {\small FCI}             & -109.2821    & 0.4022          & 2.115                   & 1.335                  & 1.94156e-4         & 2326           & 15.3 \\
      \bottomrule
    \end{tabularx}
  \end{minipage}
  \label{tab:N2_spectroscopic_params}
\end{table*}
We have to mention that the Morse potential does \emph{not}
approximate well the realistic molecular potential especially for
highly stretched configurations~\cite{Murrel:1974/jcsft2/1552}, so the
values of $D_e$ from the fit overestimate the true dissociation energy
of \ce{N2}
($D_e = \SI{228.42}{\kilo\cal\per\mol} =
\SI{0.3640}{\hartree}$~\cite{Martin:1997/jms/135}). However,
near-equilibrium properties, such as $r_e$ itself, $\omega_e$, and
$\omega_ex_e$ are accurate if fit is performed not far from the
minimum. Data in Table~\ref{tab:N2_spectroscopic_params} confirm that
$r_e$, $\omega_e$, and $\omega_ex_e$ for QCC(4)+ILCAP+BW and QCC(4)+EN
curves are very close. The root mean square (RMS) of residuals for the
former, however, is smaller than for the latter which implies the
QCC(4)+ILCAP+BW curve is smoother.

\subsection{Order-dependence of iQCC-ILCAP energies}
\label{sec:order-depend-iqcc}

The unitary~\eqref{eq:U_QCC}, which is the final form for many
\gls{VQE}-based methods, is order-dependent due to non-commutativity
of some of generators $\hat T_k$. This order-dependence has important
ramifications for state optimizations: it was demonstrated that
\emph{some} orderings may not approximate selected \gls{FCI} states
with arbitrary accuracy~\cite{Evangelista:2019/jcp/244112,
  Izmaylov:2020/pccp/12980} or may display large, of the order of
hundreds \si{\kilo\cal\per\mol}, energy errors compared to alternative
orderings~\cite{Grimsley:2020/jctc/1}.

Despite the ``ultimate'' anti-commutativity demanded by
Eq.~\eqref{eq:antisymm_cond}, the \gls{QCC}-\gls{ILCAP} Ansatz is
order-independent, which immediately follows from its equivalence to a
linear parametrization~\eqref{eq:U_ilcap}. However, some
order-dependence is brought into by the construction algorithm. On the
one hand, the reduced row-echelon form $\mathbf{M}_\text{rref}$ (see
Sec.~\ref{sec:cnot-swap-quantum}) is unique for a given matrix
$\mathbf{M}$. On the other, different ordering of columns essentially
implies that one has multiple matrices $\mathbf{M}$, each of those
leads to different anti-commutative set of Paulis. We fix the order of
columns by sorting them in accordance with descending ``importance
measure'', which is a smooth function of Hamiltonian's coefficients;
see Sec.~\ref{sec:pract-cons}. Unfortunately, some drastic changes in
composition of the anti-commutative set are still possible when
several generators acquire numerically identical measure values. These
degeneracies may be symmetry-related, if coefficients of the
Hamiltonian become equal by symmetry, or accidental. The
symmetry-related degeneracies may be fixed by introducing additional,
for example lexicographical, ordering of generators. The accidental
degeneracies are more problematic. In fact, all large kinks that are
visible in Figs.~\ref{fig:h2o_symm_pec} and \ref{fig:N2_stretch_pre}
are due to accidental degeneracies. The odds for accidental
degeneracies to occur increase with increasing the density of states
at the particular nuclear configuration and a degree of their mixing
-- that is why kinks show up with larger probabilities when chemical
bonds are ``half-broken''. Unfortunately, it is not clear how this
issue could be fixed, which warrants future studies.

\section{Conclusions}
\label{sec:conclusions}

We have presented a novel algorithm for efficient construction of
fully-anticommutative sets of Pauli generators~\eqref{eq:P_k-def} that
are tagged by additional properties, such as energy
gradients. 
We applied the Gaussian elimination procedure over the GF(2) field to
matrices that represent Pauli X~words to determine the primary and
secondary vectors (operators) in a standard basis, from which the
anti-commutative system is constructed by means of a sightly modified
Jordan--Wigner transformation. Returning from the standard to the
original system of Paulis is encoded by the matrices $\mathbf{R}^{-1}$
and $\mathbf{R}^T$ (see Sec.~\ref{sec:cnot-swap-quantum}) for Pauli X
and Z~words, respectively, and only the latter must be explicitly
computed. The algorithm complexity is linear in the size of the input
set and quadratic in the number of qubits, which allows one to apply
it to systems with hundreds of qubits and Hamiltonians containing tens
of millions terms in their Ising
decomposition~\eqref{eq:Ising_decomp}.

To demonstrate the scalability of our algorithm we have applied the
\gls{QCC}-ILCAP treatment to the symmetric stretch of the \ce{H2O}
molecule and dissociation of \ce{N2}. Both problems are considered in
reasonably large basis sets, 6-31G(d) and cc-pVDZ, respectively, which
leads to 36- and 56-qubit initial Hamiltonians containing up to
\num{e5} terms.

The \gls{QCC}-\gls{ILCAP} Ansatz was applied before (termed as
pre-processing) and after (pos-processing) ordinary \gls{iQCC}
iterations. As a pre-processing technique, the \gls{QCC}-\gls{ILCAP}
Ansatz has a limited ability to recover the correlation energy; to
account for the remaining piece we proposed the use of the
Brillouin--Wigner perturbation theory. The QCC-ILCAP+BW scheme can be
considered as a state-specific (ground-state) multiconfigurational
perturbation theory formulated in terms of qubit, rather than
fermionic, generators. Even with the perturbative correction, the
\gls{QCC}-\gls{ILCAP} treatment encounters difficulties in the case of
strong correlation which manifest themselves in non-continuous
potential energy curves. Overall, the QCC-ILCAP and QCC-ILCAP+BW
schemes are not as robust as more traditional multiconfigurational
theories, and in the present form do not allow for treatment of
excited states. However, the computational complexity is low and is
comparable to that of single-reference [\latin{e.g.} the \gls{MP2}]
perturbation theories.

The \gls{QCC}-\gls{ILCAP} treatment is most promising as a
post-processing technique. The difficulties that the
\gls{QCC}-\gls{ILCAP} treatment experiences with strongly correlated
systems are less relevant since for dressed Hamiltonians the strength
of correlation is systematically diminished. A single application of
the \gls{QCC}-\gls{ILCAP} Anzats amounts to 2--4 regular iQCC
iterations for the given qubit dimensionality (\numrange{40}{60}
qubits). For larger systems this ``efficiency ratio'' is likely to
increase because the regular \gls{QCC} Ansatz is difficult to optimize
(on a classical computer) for more than $L \sim 20$ amplitudes due to
exponential complexity, whereas the ILCAP construction and subsequent
\gls{QCC}-\gls{ILCAP} optimizations are easily done for hundreds of
them. Because of the variational nature, the \gls{QCC}-\gls{ILCAP}
energies may be taken as extrapolation to the \gls{iQCC} energies when
additional \gls{iQCC} iterations are not possible due to the excessive
size of the dressed Hamiltonian. The QCC(n)+ILCAP+BW scheme, on the
other hand, does not significantly improve upon the QCC(n)+EN
counterpart, but provides smoother potential energy curves. The energy
discrepancy between these two schemes can be used to gauge the
reliability of energy estimates. The \gls{QCC}-\gls{ILCAP} treatment
has already been used for this purpose in large-scale simulations of
\gls{OLED} materials~\cite{Genin:2022/ang/e202116175}.


\begin{acknowledgement}
  The IGR and SNG thanks Prof. Artur F. Izmaylov and Robert A. Lang
  for many fruitful discussions.
\end{acknowledgement}

\bibliography{lrgilc}

\end{document}